\documentclass[12pt]{amsart}
\usepackage{amsmath}
\usepackage{amsxtra}
\usepackage{amscd}
\usepackage{amsthm}
\usepackage{amsfonts}
\usepackage{amssymb}
\usepackage{eucal}
\usepackage{epsfig}
\usepackage{graphics}
\def\({\left(}
\def\){\right)}

\newcommand{\vphis}{\scalebox{.7}{\boldmath$\varphi$}}
\newcommand{\vphi}{\mbox{\boldmath$\varphi$}}
\newcommand{\betab}{\mbox{\boldmath$\beta$}}
\newcommand{\gammab}{\mbox{\boldmath$\gamma$}}
\newcommand{\mub}{\mbox{\boldmath$\mu$}}

\newcommand{\cb}{\mathbf{c}}
\newcommand{\bb}{\mathbf{b}}

\newcommand{\tb}{\mathbf{t}}

\newcommand{\nn}{\nonumber}
\newcommand{\bea}{\begin{eqnarray}}
\newcommand{\ena}{\end{eqnarray}}
\def\bel{\begin{eqnarray}}
\def\enl{\end{eqnarray}}
\newcommand{\be}{\begin{eqnarray*}}
\newcommand{\en}{\end{eqnarray*}}
\newcommand{\ba}{\begin{array}}
\newcommand{\ea}{\end{array}}

\newcommand{\Tr}{{\rm Tr}}

\def\[{\left[}
\def\]{\right]}
\newcommand{\la}{\lambda}

\newcommand{\al}{\alpha}

\newcommand{\z}{\zeta}

\def\half{\textstyle{\frac  1 2}}

\begin{document}

\begin{title}
{ON ONE-POINT FUNCTIONS OF DESCENDANTS IN SINE-GORDON MODEL} 
\end{title}
\author{M.~Jimbo, T.~Miwa and  F.~Smirnov}
\address{MJ: Department of Mathematics, 
Rikkyo University, Toshima-ku, Tokyo 171-8501, Japan,
E-mail: jimbomm@rikkyo.ac.jp}
\address{TM: Department of 
Mathematics, Graduate School of Science,
Kyoto University, Kyoto 606-8502, 
Japan, Email: tmiwa@math.kyoto-u.ac.jp}
\address{FS\footnote{Membre du CNRS}: Laboratoire de Physique Th{\'e}orique et
Hautes Energies, Universit{\'e} Pierre et Marie Curie,
Tour 16 1$^{\rm er}$ {\'e}tage, 4 Place Jussieu
75252 Paris Cedex 05, France, E-mail: smirnov@lpthe.jussieu.fr}

\begin{abstract}
We apply the fermionic description of 
CFT developed in our previous work to the computation of 
the one-point functions of the 
descendent fields in the sine-Gordon model.
\end{abstract}


\maketitle

\section{Introduction}
The sine-Gordon (sG) model is the most famous example of 
two-dimensional integrable Quantum Field Theory (QFT).
The sG model is defined in two-dimensional
Minkowski space  with coordinate
$\mathbf{x}=(x_0,x_1)$ by the action 
\begin{align}
\mathcal{A}_\mathrm{sG}=\int \left\{ \frac 1 {16  \pi} (\partial _\mu\vphi (\mathbf{x}))^2+\frac{2\mub ^2}{\sin(\pi\beta ^2)} \cos(\beta\vphi(\mathbf{x}))     \right\}d^2\mathbf{x}\,.
\label{action}
\end{align}
The normalisation of the dimensional coupling constant
in front of $\cos(\beta\vphi(\mathbf{x})) $  is chosen for 
future convenience.
This model has been a subject of intensive 
study during the last 30 years. 
First, by semi-classical methods 
the exact spectrum was computed, 
the factorisation of scattering was predicted, and the
exact $S$-matrix was found for certain values of the
coupling constant (in the absence of reflection of solitons) 
\cite{faddkor,neveu}.

The most significant  further results were found by the bootstrap method.
In \cite{zamS,zamzamS} the exact S-matrix was found, and in \cite{sm1986,book}
the exact form factors were computed for the energy-momentum tensor, 
topological current and the operators $e^{\pm i\beta\vphis(\mathbf{x})}$, $e^{\pm \frac{i\beta} 2 \vphis (\mathbf{x})}$.
Then the latter result was generalised to 
the operator $e^{ia\vphis(\mathbf{x})}$ 
with an arbitrary $a$ \cite{lukalpha}, 
using the methods which go back to the 
study of the closely related XXZ spin chain \cite{MJbook,MJ1996}.

It should be said that 
many important technical and conceptual methods of  
the modern theory of quantum integrable models  
originate in the quantum inverse scattering method
\cite{skl,fst}.  
It 
provides a clear mathematical interpretation
of the work by R. Baxter \cite{baxter}. 
In particular, in \cite{fst} the 
scattering matrix of the sG solitons was 
reproduced using the Bethe Ansatz. 

The knowledge of form factors 
allows us to write a series representation for the
two-point function 
\begin{align}
\langle e^{ia_1\vphis (\mathbf{x})}
e^{i a_2\vphis (0)}\rangle_\mathrm{sG}\,.
\label{corr}
\end{align}
In this paper we shall consider only the space-like 
region $\mathbf{x}^2<0$.
We shall use a lattice regularisation which breaks the
Lorentz invariance.
So we shall take 
$\mathbf{x}=(0,x)$, and use the notation $\vphi (x)=\vphi (0,x)$.
In this case 
the integrals over the form factors are rapidly convergent. 
It is rather hard to give a mathematically rigorous proof 
of the convergence of the series, 
but nobody doubts that they converge. 
Actually the convergence was proved \cite{smunpubl}
for a certain particular 
reduction of the sG model,  
known as the scaling Lee-Yang theory.  
So, the situation looks completely satisfactory. 
However, the series over the form factors converge slowly 
in the ultraviolet region (for small values of $-x^2$). 
To give an efficient description to its 
ultraviolet behaviour remained a problem 
largely open for some time. 
Before describing the way of solving
this problem let us explain 
how it is solved in a particular case.

It is known that at $\beta^2=1/2$ the sine-Gordon model is 
equivalent to the free theory of a Dirac fermion. 
The correlation functions of \eqref{corr} are non-trivial. 
The sG field $\vphi (x)$ is bilinear in fermions, 
and one has to compute 
the correlation function of two exponentials 
of bilinear forms. 
The result is obtained in \cite{SMJ}, 
generalising the seminal work \cite{WMTB} 
on the scaling Ising model. 
Namely, it is shown that the correlation function
satisfies an equivalent of 
a Painlev\'e III equation. 
The form factors are very simple in this case, 
and the form factor series 
coincides with the Fredholm determinant 
representation of this solution.
Still the problem of describing the ultra-violet
behaviour is non-trivial. 
It amounts to finding the connection coefficients 
for the Painlev\'e equation,  
which is done by studying the Riemann-Hilbert
problem \cite{Its}. 
Through this analysis, one draws an important conclusion.  
Since CFT completely describes the ultra-violet limit, 
one might na{\"{\i}}vely expect that 
the asymptotic behaviour of the two-point function
may be obtained via the perturbation theory. 
Such an assumption would imply that
the dependence 
on the mass scale is analytic in  $\mub ^2$ 
which has the dimension $(\mathrm{mass})^{2(1-\beta ^2)}$. 
However this is not the case
even for the Painlev\'e solution. 

Correct understanding of the 
conformal perturbation theory 
is one of the most important problems
in the theory of quantum integrable systems. 
This problem was studied in \cite{alzam}. 
In fact, the na{\"{\i}}ve perturbation 
theory suffers from both ultraviolet 
and infrared divergencies. 
The idea of \cite{alzam} is to absorb all 
these divergencies into non-perturbative 
data: one-point functions of primary fields 
and their descendants. 
Once it is done, 
the remaining task is a convergent 
version of conformal perturbation theory.
So the problem is divided into two steps. 
The first one requires some non-perturbative information. 
The second one is of genuinely CFT origin: actually, it
is reduced to the computation of some Dotsenko-Fateev
Coulomb gas integrals with screenings \cite{FD}.

In principle, the  procedure described in \cite{alzam} 
provides an asymptotic series in the 
ultra-violet domain which agrees with 
the structure expected 
from the Painlev\'e case: 
it is not just a power series in $\mub ^2$, but 
includes non-analytic contributions with 
fractional powers 
of $\mub ^2$. 

So the main problem is to compute the one-point
functions.
The first important result in this direction 
was achieved in \cite{lukzam}, where 
an exact formula for 
the one-point functions of the primary fields 
was conjectured. 
Then by several ingenious tricks 
(such as going form sine-Gordon to sinh-Gordon 
and back)  
a procedure was described in \cite{flzz,zzliouv,fflzz}, 
which must in principle allow us to compute 
the correlation functions of the descendants. 
Unfortunately, this procedure 
involves certain matrix Riemann-Hilbert problem 
which has  not been solved in general up to now. 
At the same time,  
this way of computation looks very indirect,
and involves steps which are hard to justify.
Still, the predicted 
result for the first not-trivial descendant \cite{fflzz} is
quite remarkable. 
Even though it was obtained by a complicated and 
non-rigorous procedure, it was checked 
against many particular cases.  
So we do not doubt in its validity. 
It will be used to check the result of the present paper. 

Let us mention here 
a deep relation between the 
sG model and the $\Phi_{1,3}$-perturbation  
of $c<1$ models of CFT \cite{BPZ}.
On the formal level the relation is simple. 
In the action \eqref{action} one can split 
$\cos (\beta\vphi(\mathbf{x}))=\half (e^{-i\beta \vphis(\mathbf{x})}+e^{i\beta \vphis(\mathbf{x})})$ and 
consider the first term as a part of the Liouville action 
and the second as a perturbation. 
The Liouville model 
with an imaginary exponent is nothing but CFT with 
the central charge
$$
c=1-6(\beta-1/\beta)^2\,,
$$
and $e^{i\beta \vphis(\mathbf{x})}$ is the field $\Phi _{1,3}(\mathbf{x})$. 
So, formally there is no 
difference between the sG model 
and the $\Phi _{1,3}$-perturbation of $c<1$ CFT. 
The situation becomes interesting 
in the case of rational $\beta ^2$. 
It was shown in \cite{sm1992,reshsm} that 
in the computation of the correlation functions of 
$e^{\frac{in\beta} 2\vphis (x)}$ ($n=1,2,3,\cdots$), 
a restriction of degrees of freedom 
takes place for the intermediate states of solitons. 
The mechanism of reduction 
is similar to the RSOS restriction for vertex models. 
This phenomenon is known as the 
restriction of the sG model.
For example, if $\beta^2=3/4$ 
the solitons reduce to Majorana fermions, and
the restricted model is nothing but the scaling Ising model.

In the series of papers, \cite{HGS,HGSII,HGSIII,HGSIV}
which we will refer to as I,II,III,IV, respectively,
we studied the hidden fermionic structure of the XXZ spin chain. 
In particular, in IV
the relation between our fermions in the scaling
limit and $c<1$ CFT was established. 
The long distance behaviour 
of the XXZ model and the short 
distance behaviour of the sG model are described by the same CFT:  
free bosons with the compactification radius $\beta^2$.  
For the XXZ model, we use
the coupling parameter 
$\nu$ 
related to $\beta^2$ via 
$$\beta^2=1-\nu\,.$$
This identification is used when relating 
the results of IV to those of \cite{BLZI,BLZII},  
which were important for us. 
The relevant CFT has the central charge
$$
c=1-6\frac {\nu^2}{1-\nu}\,.
$$
Here we do not consider the peculiarity of a 
rational $\nu$, so using the $c<1$ CFT means
just the usual modification of the energy-momentum tensor. 
In the ultraviolet limit, the  
sG model is described by two chiral copies of CFT.  
We use the notation
$$ 
\Phi _\al(x)=e^{
\ \frac { \nu}{1-\nu}\al 
\left\{\frac {i\beta} 2\vphis (x)\right\}}\,.
$$
The field
$i\beta\vphi(x)$
splits into two chiral fields 
$2\varphi(x)+2\bar{\varphi}(x)$.
(We follow the normalisation of the fields $\varphi(x),\bar{\varphi}(x)$
given in \cite{BLZI,BLZII} and in IV.)
Our goal is to compute the one point functions
$$\frac{\langle P(\{\mathbf{l}_{-m}\})\bar{P}(\{\bar{\mathbf{l}}_{-m}\})\Phi _\al (0)\rangle_{sG}}
{\langle \Phi _\al(0)\rangle_{sG}}\,,$$
where $P(\{\mathbf{l}_{-m}\})$, 
$\bar{P}(\{\bar{\mathbf{l}}_{-m}\})$ are polynomials in
the generators of two chiral copies of the Virasoro algebra with the central charge $c$.

The universal enveloping algebra of 
the Virasoro algebra contains the local
integrals of motion $\mathbf{i}_{2k-1}$ 
($\mathbf{i}_{1}=\mathbf{l}_{-1}$) \cite{zamint}
which survive the $\Phi _{1,3}$-perturbation. 
Clearly, the one-point functions of 
descendants created by them vanish. 
We assume that it is possible to write any element of the 
chiral Verma module generated by
$\Phi_\al(0)$
as 
$P_1(\{\mathbf{i}_{2k-1}\})
P_2(\{\mathbf{l}_{-2k}\})\Phi _\al(0)$. 
So, actually we are interested in computing
$$
\frac{\langle P(\{\mathbf{l}_{-2m}\})
\bar{P}(\{\bar{\mathbf{l}}_{-2m}\})\Phi _\al (0)\rangle_{sG}}
{\langle \Phi _\al(0)\rangle_{sG}}\,.
$$
For the simplest non-trivial 
case $\mathbf{l}_{-2}\bar{\mathbf{l}}_{-2}\Phi _\al (0)$ 
the problem was solved in \cite{fflzz}. If one wants to consider the descendants
created by the Heisenberg algebra it is easy to do using the formulae
$$(1-\nu) T(z)=:\varphi '(z)^2: +\nu\varphi (z),\quad
(1-\nu) \bar T(\bar z)=:\bar\varphi '(\bar z)^2: +\nu\bar\varphi (\bar z)\,.$$
In this paper we shall consider only the domain $0<\al<2$, but the final results
allow analytical continuation for all values of $\al$.

The whole idea of IV is that 
the usual basis of the Verma module is not
suitable for the perturbation, 
and we have to introduce another one. 
For the moment we consider only one chirality. Working
modulo action of the local integrals of motion the new basis is provided by
uncharged products of two fermions $\betab ^{\mathrm{CFT}*}_{2k-1}$ and
$\gammab ^{\mathrm{CFT}*}_{2k-1}$. 
The fermions respect the Virasoro grading:
$$[\mathbf{l}_0 ,  \betab ^{\mathrm{CFT}*}_{2j-1}\gammab ^{\mathrm{CFT}*}_{2k-1}]
=(2j+2k-2)\betab ^{\mathrm{CFT}*}_{2j-1}\gammab ^{\mathrm{CFT}*}_{2k-1}\,.$$
We have 
\begin{align}
\betab ^{\mathrm{CFT}*}_{I^+}\gammab ^{\mathrm{CFT}*}_{I^-}\Phi _\al (0)=
\left\{P_{I^+,I^-}^{\mathrm{even}}(\{  \mathbf{l}_{-2k}  \})+d_\al P
_{I^+,I^-}^{\mathrm{odd}}(\{  \mathbf{l}_{-2k}  \})\right\}
\Phi _\al (0)\,,\label{transform}
\end{align}
where $I^{+}$ and $I^-$ are ordered multi-indices:
for $I=(2r_1-1,\cdots 2r_n-1)$ with $r_1<\cdots <r_n$, we set 
$$
\betab^{\mathrm{CFT}*}_{I}=\betab ^{\mathrm{CFT}*}_{2r_1-1}
\cdots \betab ^{\mathrm{CFT}*}_{2r_n-1}\,,
\quad 
\gammab ^{\mathrm{CFT}*}_{I}=\gammab ^{\mathrm{CFT}*}_{2r_n-1}
\cdots \gammab ^{\mathrm{CFT}*}_{2r_1-1}\,,
$$
and similarly for $\bar{\betab}_I$, $\bar{\gammab}_I$.
We require $\#(I^+)=\#(I^-)$.  
In the right hand side of \eqref{transform},
$P_{I^+,I^-}^{\mathrm{even}}(\{  \mathbf{l}_{-2k}  \})$ and
$P_{I^+,I^-}^{\mathrm{odd}}(\{  \mathbf{l}_{-2k}  \})$ are homogeneous polynomials,
the constant $d_\al$ is given by
$$d_\al =\frac{\nu(\nu-2)}{\nu-1}(\al-1)= \textstyle{\frac 1 6}\sqrt{(25-c)(24\Delta _\al +1-c)}\,,$$
and the separation into even and odd parts is determined by the reflection 
 $\betab ^{\mathrm{CFT}*}_{I^+}\gammab ^{\mathrm{CFT}*}_{I^-}\ \leftrightarrow
\ \betab ^{\mathrm{CFT}*}_{I^-}\gammab ^{\mathrm{CFT}*}_{I^+}$. 

The coefficients of
the polynomials $P_{I^+,I^-}^{\mathrm{even}}$ and $P_{I^+,I^-}^{\mathrm{odd}}$
are rational functions of $c$ and 
$$\Delta _{\al}=\frac {\al (\al-2)\nu ^2}{4(1-\nu)}$$ 
only. The denominators 
factorise into multipliers $\Delta _{\al}+2k$, $k=0,1,2,\cdots$. Exact formulae up 
to the level 6 can be found in IV, Eq.~(12.4). In particular, on the level 2 we have
$P_{1,1}^\mathrm{even}=1$, $P_{1,1}^\mathrm{odd}=0$. 
The transformation
\eqref{transform} is invertible.

The operators $\betab ^{\mathrm{CFT}*}_{2k-1}$ and
$\gammab ^{\mathrm{CFT}*}_{2k-1}$ were found as the scaling limit of the fermions which
create the quasi-local fields for the XXZ spin chain. This became possible 
after the computation of the expectation values of the quasi-local
fields on the cylinder (see III). Actually, two different scaling limits are
possible, and the second one provides the fermionic operators for the second
chirality: $\bar{\betab} ^{\mathrm{CFT}*}_{2k-1}$ and
$\bar{\gammab }^{\mathrm{CFT}*}_{2k-1}$. The same story repeats for these operators, in particular,
we have 
\begin{align}
\bar{\betab} ^{\mathrm{CFT}*}_{\bar{I}^+}\bar{\gammab} ^{\mathrm{CFT}*}_{\bar{I}^-}\Phi _\al (0)=
\left\{P_{\bar{I}^+,\bar{I}^-}^{\mathrm{even}}(\{  \bar{\mathbf{l}}_{-2k}  \})-d_{\al} P
_{\bar{I}^+,\bar{I}^-}^{\mathrm{odd}}(\{  \bar{\mathbf{l}}_{-2k}  \})\right\}
\Phi _\al (0)\,.\label{transformbar}
\end{align}

The main statement of this paper is that in the fermionic basis 
the sG one-point
functions are simple:
\begin{align}
&\frac
{\langle  
\bar{\betab} ^{\mathrm{CFT}*}_{\bar{I}^+}\bar{\gammab} ^{\mathrm{CFT}*}_{\bar{I}^-}
{\betab} ^{\mathrm{CFT}*}_{{I}^+}{\gammab} ^{\mathrm{CFT}*}_{{I}^-}\Phi _\al (0)
\rangle_\mathrm{sG}}
{\langle  
\Phi _\al (0)
\rangle_\mathrm{sG}}=
(-1)^{\#(I^+)}\delta _{\bar{I}^-,I^+}\delta _{\bar{I}^+,I^-}
\label{main}\\
&\times \(
\frac
{ M  \   \sqrt{\pi}   \  \Gamma(\frac 1 {2\nu})  }
{   2      \sqrt{1-\nu} \  \Gamma(\frac {1-\nu} {2\nu})}
\)^{2|I^+|+2|I^-|}\prod _{2n-1\in I^+}G_{n}(\al)
\prod _{2n-1\in I^-}G_{n}(2-\al)\,,
\nn
\end{align}
Here 
$$
G_{n}(\al)=(-1)^{n-1}((n-1)!)^2
\frac {\Gamma \(\frac {\al}2 +\frac {1-\nu}{2\nu}(2n-1)\)
\Gamma \(1-\frac {\al}2 -\frac {1}{2\nu}(2n-1)\)}
{\Gamma \(1-\frac {\al}2 -\frac {1-\nu}{2\nu}(2n-1)\)
\Gamma \(\frac {\al}2 +\frac {1}{2\nu}(2n-1)\)}\,,
$$
$|I|$ stands for the sum of elements of $I$, and $M$ is the mass of soliton. 
We recall that  $\#(I^+)=\#(I^-)$ is required  in order
to stay in the same Verma module, 
and $\#(\bar{I}^+)=\#(\bar{I}^-)$ follows.

Using \eqref{main} we can find all the one-point functions of
descendants. Up to the level 6 the results of IV can be used. In particular,
at the level 2 we find a perfect agreement with the formula
(1.8) of
\cite{fflzz} 
after the identification: $\eta =\al -1$, $\xi=\frac {1-\nu}{\nu}$.
To proceed to levels higher than 6 one should
perform further computations in the spirit of IV.

Let us explain how we proceed in justification of the main formula \eqref{main}.
For CFT we use the lattice regularisation by the six vertex model
(equivalently XXZ spin chain). 
For this model we use the fermionic 
description of the space of quasi-local operators 
found in I,II.
On the lattice we have creation  operators
$\bb ^*(\z), \cb^*(\z)$ and 
annihilation operators
$\bb (\z), \cb(\z)$. 
The most honest way to proceed to the sG model would
be to regularise it via the eight vertex model, 
and then to consider the scaling limit. 
On this way we would meet two problems.
The first is the $U(1)$ symmetry which 
is broken in the eight vertex model. 
It is hard to introduce the lattice analogue of $\Phi _{\al}$ 
with arbitrary $\al$. 
This difficulty may be overcome by 
going to the SOS model. 
The second problem is conceptually more difficult. 
For the elliptic $R$-matrices we 
do not have an analogue of the $q$-oscillators 
which is crucial for the construction 
of our fermionic operators. 
For the moment we do not know how to attack this problem. 
Let us notice, however, that in the case $\al=0$ 
an analogue of bilinear combinations of the
annihilation operators $\bb(\z)\cb(\xi)$ exists. 
It is defined in the papers \cite{XYZ,Algebraic}.

So, having problems with the eight vertex model 
we are forced to take another approach. 
In the paper \cite{DDV} the sG or the massive Thirring
models was obtained as a limit of the inhomogeneous 
six vertex model 
with the inhomogeneity $\z _0$ (see section \ref{sG} below).
Notice that, for this limit to make sense,  
one has first to consider 
the finite lattice on $\mathbf{n}$ sites, 
and then take the 
limit $\z _0\to \infty$,  $\mathbf{n}\to \infty$ 
in a concerted way in order that the finite mass scale appears. 
But exactly this kind of procedure became very natural
for us after we had computed in III the expectation
values of quasi-local operators on the cylinder. 
The compact direction on the cylinder 
is called the Matsubara direction, and its size
$\mathbf{n}$ is what is needed for 
considering the limit in the spirit of \cite{DDV}.

The paper is organised as follows. 
In Section \ref{XXZ} we review our previous paper IV, 
and explain how to obtain the fermionic 
description for two chiral CFT models 
from the XXZ spin chain (six vertex model).
In Section \ref{sG} we introduce the  
inhomogeneous six vertex model 
and consider the 
continuous limit which produces 
the sG model according to \cite{DDV}.
We derive the one-point functions 
using the fermionic description of  
ultra-violet CFT.
\section{Two scaling limits of the XXZ model and 
two chiralities}\label{XXZ}

Our study of the XXZ model is based on the fermionic
operators defined in I,II. This definition allowed
us to compute in III the following expectation
value. 
Consider a homogeneous six vertex model 
on an infinite cylinder. 
Let $T_\mathrm{S,M}$ be the monodromy matrix, 
where $\rm S$
refers to the 
infinite direction (called the space direction), 
and $\rm M$
refers to the compact circular direction 
(called the Matsubara direction). 
We use  $\mathbf{n}$ to denote the length of the latter. 
We follow the notations in IV.
For a quasi-local operator $q^{2\al S(0)}\mathcal{O}$ 
on the spacial lattice, 
we consider 
\begin{align}
Z_{\mathbf{n}}^{\kappa}
\Bigl\{q^{2\al S(0)}\mathcal{O}
\Bigr\}=
\frac{\Tr _{\mathrm{S}}
\Tr_{\mathrm{M}}\Bigl(T_{\mathrm{S},\mathrm{M}}
q^{2\kappa S+2\al S(0)}\mathcal{O}\Bigr)}
{\Tr _{\mathrm{S}}\Tr _\mathrm{M}\Bigl(T_{\mathrm{S},\mathrm{M}}q^{2\kappa S+2\al S(0)}\Bigr)}
\,.
\label{Zkappa}
\end{align}
The generalisation of this functional 
$Z_{\mathbf{n}}^{\kappa,s}$
was introduced in IV. In the scaling limit, the introduction of $s$ in this functional
amounts to changing (screening) the background charge at $x=-\infty$
by $-2s\frac{1-\nu}\nu$. It enables us to deal with the special case of
the functional for which the effective action of local integrals of motion becomes trivial.

The quasi-local operators are created from the primary field $q^{2\al S(0)}$ by action of
the creation operators $\tb ^*(\z), \bb ^*(\z), \cb ^*(\z)$. Actually they act on the space
$$
\mathcal{W}^{(\al)}
=\bigoplus\limits _{s=-\infty}^{\infty}\mathcal{W}_{\al-s,s}
\,,
$$
where $\mathcal{W}_{\al-s,s}$ denotes the space of quasi-local operators of spin $s$ with tail $\al -s$. 

In this paper we shall consider the subspace $\mathcal{W}^{(\al)}_\mathrm{ferm}$ 
of the space $\mathcal{W}^{(\al)}$ which are created from the primary fields 
only by fermions $\bb ^*_p$, $\cb^*_p$ (see \eqref{b*} below). On 
$\mathcal{W}^{(\al)}_\mathrm{ferm}$ 
 the functional 
$Z_\mathbf{n}^{\kappa ,s}$ 
allows the 
determinant form which is convenient to summarise as
$$
Z_\mathbf{n}^{\kappa,s}\bigl\{q^{2\al S(0)}
\mathcal{O}\bigr\}=\frac {\Tr _\mathrm{S}\(e^{\Omega_\mathbf{n}}
\bigr( q^{2\al S(0)}
\mathcal{O}  \bigl)\)}
{\Tr _\mathrm{S}\( q^{2\al S(0)}  \)}\,,
$$
where 
$$
\Omega _\mathbf{n}=
\frac{1}{(2\pi i)^2}
\int\limits _\Gamma\!\!\int\limits _\Gamma \omega_{\mathrm{rat},\mathbf{n} }(\z,\xi)
\cb (\xi)\bb (\z)\frac{d\z^2}{\z ^2}\ \frac{d\xi^2}{\xi ^2}\,,
$$
where the contour $\Gamma$ goes around $\z^2=1$. 

The function $\omega_{\mathrm{rat},
\mathbf{n} 
}(\z,\xi)$ is defined by the Matsubara data
(see III, \cite{BG}, IV).
Besides the length of the Matsubara chain $\mathbf{n}$
it depends on the parameters $\kappa$,
$\al$, $s$ and on possible inhomogenieties in  the Matsubara chain. 
However, we mark explicitly only the 
dependence on $\mathbf{n}$
which is the most important for us here. 

We have ignored the descendants created by $\tb ^*(\z)$. 
Actually, they do not give any non-trivial contributions  
in the limit $\mathbf{n}\to \infty$, 
which we shall be interested in. 
Then $Z^{\kappa}_\infty$ is automatically reduced 
to the the quotient space:
$$
\mathcal{W}^{(\al)}_\mathrm{quo}
=\mathcal{W}^{(\al)}\scalebox{1.1}
{/}(\tb ^*(\z)-2)\mathcal{W}^{(\al)}\,,
$$
which is obviously isomorphic to $\mathcal{W}^{(\al)}_\mathrm{\mathrm ferm}$ as a linear space. 

We shall not repeat the definition of 
$\omega_{\mathrm{rat},
\mathbf{n} 
}(\z,\xi)$
because 
in the present paper we shall use only very limited information
about it.
Let us explain, however, the suffix ``rat": the function 
$(\xi/\z)^{\al}\omega_{\mathrm{rat},
\mathbf{n} 
}(\z,\xi)$
is a rational function of $\z^2$ and $\xi ^2$. 
The really interesting situation occurs when $\mathbf{n}\to\infty$.
In that case the Bethe roots for the transfer matrices in the Matsubara
direction become dense on the half axis $\z^2>0$. If we do not
introduce additional rescaling as described below, 
the function
$\omega_{\mathrm{rat},
\mathbf{n} 
}(\z,\xi)$ goes to the simple limit:
\begin{align}
\omega_{\mathrm{rat},
\mathbf{n} 
}(\z,\xi)\ \ \ \longrightarrow
\hskip -0.8cm{}_{{}_{{}_{\scalebox{.7}{$\mathbf{n}\to\infty$}}}}\ 4
\omega_{0} (\z/\xi,\al)+\nabla\omega  (\z/\xi,\al)
\,,\label{limomega}
\end{align}
where 
\begin{align}
&4\omega_{0} (\z,\al)=-\int\limits
_{-i\infty}^ {i\infty}\z ^{u}\frac {\sin\frac {\pi} 2((1-\nu)u-\al)}
{\sin\frac {\pi} 2(u-\al) \cos\frac {\pi\nu } 2u}du\,,\label{omega}\\
&\nabla\omega  (\z,\al)=-\psi (\z q,\al)+\psi (\z q^{-1},\al)+2i\z^{\al}\tan\Bigl(\frac {\pi\nu\al} 2\Bigr)\,,\nn\\
&\psi (\z,\al)=\z ^{\al}\frac {\z^2+1}{2(\z ^2-1)}\,.\nn
\end{align}
The reason for extracting the elementary $\nabla\omega $ is 
due to the fact that the function $\omega_{ 0} $
satisfies the relation typical for CFT 
\begin{align}
&\omega_{0} (\z ,\al)=\omega_{0} (\z ^{-1},2-\al)\,.\label{symom}
\end{align}
Notice that $(\xi/\z)^{\al}\omega_{0} (\z/\xi,\al)$ is not a single-valued function of
$\z^2$ and $\xi ^2$. So, the property of rationality is lost in the limit.
We define
 $\Omega_{0} $ and $\nabla\Omega $ in the same way as $\Omega _{\mathbf{n}}$
 replacing $\omega _{\mathrm{rat},\mathbf{n}}(\z,\xi)$ by 
 respectively $4\omega_{0} (\z/\xi,\al)$ and $\nabla\omega (\z/\xi,\al)$.

Following IV
we denote  the original creation operators introduced in II
by $\bb_\mathrm{rat} ^*(\z)$ and $\cb_\mathrm{rat} ^*(\z)$.
They satisfy the property:
$$\Tr_\mathrm{S}\(\bb_\mathrm{rat} ^*(\z)(X)\)=0,\quad
\Tr_\mathrm{S}\(\cb_\mathrm{rat} ^*(\z)(X)\)=0\,,$$
for all quasi-local operators $X$.
In the present paper it is useful to
replace these operators by the following Bogolubov transformed ones:
\begin{align}
\bb _{0}^*(\z)=e^{-\nabla\Omega }\bb_\mathrm{rat} ^*(\z)
e^{\nabla\Omega},\quad 
\cb _{0}^*(\z)=e^{-\nabla\Omega}\cb_\mathrm{rat} ^*(\z)
e^{\nabla\Omega}\,.\label{bogolub}
\end{align}
Obviously, the functional $Z^{\kappa,s}_\mathbf{n}$ calculated on the descendants
generated by these operators is expressed as determinant constructed from function
$\omega _{\mathrm{rat},\mathbf{n}}(\z,\xi)-\nabla\omega (\z/\xi,\al)$ which in the limit
$\mathbf{n}\to\infty$ goes to $4\omega_{0}(\z/\xi,\al)$. 

As it is explained in IV,
the operators $\z ^{-\al}\bb _{0}^*(\z)$ and $\z ^{\al}\cb _{0}^*(\z)$ are rational functions of $\z^2$
as far as they are considered in the functional $Z^{\kappa,s}_{\mathbf{n}}$, and
they have the following behaviour at $\z ^2=0$:
$$\bb ^*_{0}(\z)=\sum\limits_{j=1}^{\infty} \z ^{\al +2j-2}\bb ^*_{\mathrm{screen},j},
\quad \cb ^*_{0}(\z)=\sum\limits_{j=1}^{\infty} \z ^{-\al +2j}\cb ^*_{\mathrm{screen},j}\,.$$
For $\bb ^*_{\mathrm{screen},j}$, 
$\cb ^*_{\mathrm{screen},j}$we have used the suffix ``screen" to stand for ``screening" in view of its similarity to
the lattice screening operators used in IV.


In IV, another set of operators $\bb^*(\z)$, $\cb^*(\z)$ is obtained from $\bb^*_{\rm rat}(\z)$, $\cb^*_{\rm rat}(\z)$
by a kind of Bogolubov transformation which contains $\tb^*(\z)$.
We shall not write explicitly this Bogolubov
transformation, but only the one relating $\bb^*(\z)$, $\cb ^*(\z)$
to $\bb_0^*(\z)$, $\cb _0^*(\z)$, both acting
on the quotient space $\mathcal{W}^{(\al)}_\mathrm{quo}$ because only these operators are
used in this paper. 
The operators $\bb^*(\z)$, $\cb ^*(\z)$ are important because 
$Z^{\kappa,s}_\infty $ vanishes on their
descendants. Notice that we do not allow $\kappa,s$
to grow together
with $\mathbf{n}$. In that case the dependence on $\kappa,s$ disappears for $\mathbf{n}=\infty$.


\vskip.3cm\noindent
{\bf Remark.} The defining equations for 
$\omega _{\mathrm{rat}, \mathbf{n}}(\z,\xi)$ given in
IV imply that 
the limit \eqref{limomega} has a very general nature. 
Namely, 
the result is independent 
not only of $\kappa, s$ but also of 
inhomogeneities in the Matsubara chain.
The situation is similar to that for the $S$-matrix  
which is the same for homogeneous, inhomogeneous 
XXZ chains or even for the sG model.

\vskip.3cm

If we consider the Bogolubov transformation which connects the operators $\bb^*(\z)$, $\cb^*(\z)$ and
$\bb^*_{0}(\z)$, $\cb^*_{0}(\z)$ as acting on the quotient space $\mathcal{W}^{(\al)}_\mathrm{quo}$,
it reduces to
\begin{align}
\bb^*(\z)=e^{-\Omega _0}\bb^*_{0}(\z)e^{\Omega _0},\quad
\cb^*(\z)=e^{-\Omega _0}\cb^*_{0}(\z)e^{\Omega_0}\,.
\label{bb0}
\end{align}

%
%
We catch operators acting on $\mathcal W^{(\al)}$ by developing $\bb^*(\z)$, $\cb^*(\z)$ and
$\bb(\z)$, $\cb(\z)$ around the point $\z^2=1$:
\begin{align}
&\bb ^*(\z)\ \ \simeq\hskip -.6cm{}_{{}_{{}_{\scalebox{.7}{$\z^2\to 1$}}}}
\ \ 
\sum\limits_{p=1}^{\infty} (\z ^2-1)^{p-1}\bb^*_p, \qquad \cb ^*(\z)
\ \ \simeq\hskip -.6cm{}_{{}_{{}_{\scalebox{.7}{$\z^2\to 1$}}}}
\ \ 
\sum\limits_{p=1}^{\infty} (\z ^2-1)^{p-1}\cb^*_p\,,\label{b*}\\
&\bb (\z)=\sum\limits_{p=0}^{\infty} (\z ^2-1)^{-p}\bb_p, \qquad \cb (\z)=
\sum\limits_{p=0}^{\infty} (\z ^2-1)^{-p}\cb_p\,.
\end{align}
These operators are (quasi-)local in the sense of II, Section 3.3,
while $\bb ^*_{\mathrm{screen},j}$ and $\cb ^*_{\mathrm{screen},j}$
are highly ``non-local".

The operators $\bb^*_p$, $\cb^*_p$ create quasi-local operators in the sense of
II, Section 3.1 by acting on the primary field.
Here we slightly change the notation compared to II,III,
where the Fourier coefficients are defined after 
removing an overall power $\z ^{\pm \al}$. 
The reason was that we wanted that the result of action
of $\bb^*_p$, $\cb^*_p$ is a rational function of $q$, $q^\al$. 
This rationality property
is irrelevant in this paper, 
and extracting $\z ^{\pm \al}$ may even cause a confusion.

So far we have been discussing the simple limit $\mathbf{n}\rightarrow\infty$.
Now we discuss the scaling limit to CFT.
In IV we studied the  scaling limit of the 
homogeneous XXZ chain on the cylinder.
The key idea is to consider first of all the scaling limit in the Matsubara direction. 
Namely, denoting  the length of the Matsubara chain
by  $\mathbf{n}$ and introducing the step of the lattice $a$ we consider the limit 
\begin{align}
\mathbf{n}\to\infty,\quad a\to 0,\quad 
\mathbf{n}a=2\pi R \ \ \mathrm{fixed}\,. 
\label{scal}
\end{align}
The requirement is that 
if we rescale the spectral parameter as $\z = (Ca)^{\nu}\la$,
the Bethe roots for the transfer matrix in the Matsubara direction which are close to $\z^2=0$
remain finite.
The constant $C$ is chosen to have an agreement with CFT as
\begin{align}
C=\  \frac{\Gamma\(\frac {1-\nu}{2\nu}\)}{2\sqrt{\pi}\ \Gamma \(\frac 1{2\nu}\)}\Gamma (\nu)^{\frac 1\nu}\,.
\label{C}
\end{align}

Next we consider the scaling limit in the space direction.
We conjecture that under the presence of the background charges effected by
the screening operators, the lattice operator $q^{2\kappa S+2\al S(0)}$ goes to the limit
$\Phi_{1-\kappa'}(-\infty)\Phi _\al (0)\Phi_{1+\kappa}(\infty)$
where $$\kappa'=\kappa+\al+2s\frac{1-\nu}\nu.$$
Furthermore,
we have shown that in the weak sense the following limits exist
\begin{align}
&2\betab ^*(\la)
=\lim_{a\to 0}\bb ^*((Ca)^{\nu}\la)  ,\quad
2\gammab ^*(\la)=\lim_{a\to 0}\cb ^*((Ca)^{\nu}\la)\,.
\label{limbc}
\end{align}

The operators $\betab^*(\la)$, $\gammab^*(\la)$ 
have the asymptotics
at $\la ^2\to \infty$:
\begin{align}
\betab^*(\la)\simeq\sum\limits _{j=1}^{\infty}
\la ^{-\frac{2j-1}\nu}\betab^*_{2j-1},\quad
\gammab^*(\la)\simeq\sum\limits _{j=1}^{\infty}
\la ^{-\frac{2j-1}\nu}\gammab^*_{2j-1}\,,
\label{betgam}
\end{align}
where $\betab^*_{2j-1}$, $\gammab^*_{2j-1}$  act between
different Verma modules as follows
\begin{align}
&\betab ^*_{2j-1}\ \ :\  
\mathcal{V}_{\al+2\frac{1-\nu}{\nu}(s-1)}
\otimes \overline{\mathcal{V}}_{\al} \ 
\to\  
\mathcal{V}_{\al+2\frac{1-\nu}{\nu}s}
\otimes \overline{\mathcal{V}}_{\al}\,,
\nn\\ 
&\gammab^*_{2j-1}\ \ :\  
\mathcal{V}_{\al+2\frac{1-\nu}{\nu}(s+1)} 
\otimes \overline{\mathcal{V}}_{\al}\ 
\to\  
\mathcal{V}_{\al+2\frac{1-\nu}{\nu}s}
\otimes \overline{\mathcal{V}}_{\al}\,.
\nn
\end{align}
The action on the second component is trivial.
The functional $Z^{\kappa,s}_\mathbf{n}$ turns in this limit
into the three-point function for $c<1$ CFT as explained in
IV. The identification of 
descendants created by $\betab^*$ 
and $\gammab^*$ with Virasoro descendants 
is made by studying the function
$$
{\omega }_R(\la,\mu)
=\frac14(\lim_{\mathrm{scaling}}\omega _{\mathrm{rat},\mathbf{n}}((Ca)^{\nu}\la, 
(Ca)^{\nu}\mu)
-\nabla\omega(\la/\mu,\al))\,,
$$
where $\lim_{\mathrm{scaling}}$ refers to the scaling limit 
\eqref{scal}. 
Contrary to the simple-minded limit 
$\mathbf{n}\to \infty$ \eqref{limomega}, 
the function $(\mu/\la)^{\al}{\omega }_R(\la,\mu)$ 
remains a single-valued function of $\la ^2$ and $\mu^2$, 
but it develops an essential singularity at $\la^2=\infty$,
$\mu ^2=\infty$.

In the present paper we shall consider sG model which requires
putting together  the two chiralities.  To this end
we shall need to consider
not only the asymptotical region $\la^2\to\infty$, but also
$\la ^2\to 0$. Analysing the function ${\omega }_R(\la,\mu)$ one concludes that
the following limits exist
\begin{align}
&\half\lim\limits_{a\to 0}\bb _{0}^*((Ca)^{\nu}\la)
\ \ \simeq\hskip -.6cm{}_{{}_{{}_{\scalebox{.6}
{$\la^2\to 0$}}}}
\ \betab_\mathrm{screen} ^*(\la)=\sum\limits _{j=1}^{\infty}\la ^{\al+2j-2}
\betab_{\mathrm{screen},j}^*\,,\label{scr1}\\
&\half\lim\limits_{a\to 0}\cb _{0}^*((Ca)^{\nu}\la)
\ \ \ \simeq\hskip -.6cm{}_{{}_{{}_{\scalebox{.6}
{$\la^2\to 0$}}}}
\  \gammab_\mathrm{screen} ^*(\la)=\sum\limits _{j=1}^{\infty}\la ^{-\al+2j}
\gammab_{\mathrm{screen},j} ^*\,.\nn
\end{align}
For the moment we do not know how to use these operators, but
one thing is clear: they create highly non-local fields. 

The $L$-operator depends on $\z$ and $\z^{-1}$ in
a  symmetric way. 
That is why  another scaling limit is possible:
\begin{align}
&2\bar{\betab }^*(\la)
=\lim_{a\to 0}\bb ^*((Ca)^{-\nu}\la)  ,\quad
2\bar{\gammab} ^*(\la)=\lim_{a\to 0}\cb ^*((Ca)^{-\nu}\la)\,,
\label{deftbgbar}
\end{align}
which allow the power series at $\la\to 0$:
\begin{align}
\bar\betab^*(\la)\simeq\sum\limits _{j=1}^{\infty}
\la ^{\frac{2j-1}\nu}\bar\betab^*_{2j-1},\quad
\bar\gammab^*(\la)\simeq\sum\limits _{j=1}^{\infty}
\la ^{\frac{2j-1}\nu}\bar\gammab^*_{2j-1}\,.
\label{betgambar}
\end{align}
The resulting operators act as follows
\begin{align}
&\bar{\betab} ^*(\la)\ \ :\  
\mathcal{V}_{\al}
\otimes \overline{\mathcal{V}}_{\al
 +2\frac{1-\nu}{\nu}(s-1)  } \ 
\to\  
\mathcal{V}_{\al}
\otimes \overline{\mathcal{V}}_{\al
   +2\frac{1-\nu}{\nu}s }\,,
\nn\\ 
&\bar{\gammab}^*(\la)\ \ :\  
\mathcal{V}_{\al}
\otimes \overline{\mathcal{V}}_{\al
 +2\frac{1-\nu}{\nu}(s+1)}      \ 
\to\  
\mathcal{V}_{\al}
\otimes \overline{\mathcal{V}}_{\al
   +2\frac{1-\nu}{\nu}s}   \,.
\nn
\end{align}
These operators are obtainable from the previous ones
by the substitution
\begin{align*}
\{\al,\ \la \}\ \to
\{2-\al,\ \la ^{-1}\}\,.
\end{align*}

The proof goes through 
considering the 
scaling 
limit  \eqref{scal} 
and studying the function
$$
\bar{\omega }_R(\la,\mu)=
\frac14(\lim _{\mathrm{scaling}}\omega 
_{\mathrm{rat},\mathbf{n}}((Ca)^{-\nu}\la, 
(Ca)^{-\nu}\mu)-\nabla\omega(\la/\mu,\al))\,.
$$

The analysis is parallel to
the one performed for the first chirality in IV.
The function $(\mu/\la)^{\al}\bar{\omega }_R(\la,\mu)$ is a single-valued
function of $\la^2$, $\mu ^2$ with essential singularities at
$\la ^2=0$, $\mu^2=0$.

The operators $\bar{\betab}^*_{\mathrm{screen}}(\la)$ and $\bar{\gammab}^*_{\mathrm{screen}}(\la)$ are introduced similarly to the first chirality.
\begin{align}
&\half\lim\limits_{a\to 0}\bb _{0}^*((Ca)^{-\nu}\la)\ \ \simeq\hskip -.6cm{}_{{}_{{}_{\scalebox{.6}
{$\la^2\to \infty$}}}} 
\ \bar{\betab}_\mathrm{screen} ^*(\la)=
\sum\limits_{j=1}^{\infty}\la ^{\al-2j}\bar{\betab}^*_{
\mathrm{screen},j}\,,\label{screenbar}\\
&\half\lim\limits_{a\to 0}\cb _{0}^*((Ca)^{-\nu}\la)\ \ \ \simeq\hskip -.6cm{}_{{}_{{}_{\scalebox{.6}
{$\la^2\to \infty$}}}}
\ \bar{\gammab}_\mathrm{screen} ^*(\la)=
\sum\limits_{j=1}^{\infty}\la ^{2-\al-2j}\bar{\gammab}^*_{
\mathrm{screen},j}\,.\,.\nn
\end{align}

For both chiralities we have
$$\omega _R(\la,\mu)
\ \ \longrightarrow
\hskip -.8cm{}_{{}_{{}_{\scalebox{.7}{$R\to\infty$}}}}\ 
 \omega_{0} (\la/\mu, \al),\quad 
\bar{\omega }_R(\la,\mu) 
\ \ \longrightarrow
\hskip -.8cm{}_{{}_{{}_{\scalebox{.7}{$R\to\infty$}}}}\ 
\omega_{0}(\la/\mu,\al)\,.$$
So, the na{\"{\i}}ve $\mathbf{n}\to \infty$ limit is reproduced. 


\section{Inhomogeneous six vertex model 
and sine-Gordon model}\label{sG}

We want to put two chiral models together and to make them interacting.
According to the previous discussion the lattice analogue
of chiral CFT is the XXZ model. 
So, the two non-interacting chiral models correspond 
to the lattice containing 
two non-interacting six vertex sublattices. 
As an 
example we shall consider 
the ``even'' or ``odd'' sublattices, consisting of lattice points 
with coordinates $(j,\mathbf{m})$ such that 
$j,\mathbf{m}$ are both even or both odd. 
It is well-known \cite{DDV} how to force these two to interact:
one has to consider a six vertex model
on the entire lattice
with alternating inhomogeneneity parameters. 
We denote $\mathbf{S}=\mathrm{S}\cup
\overline{\mathrm{S}}$, $\mathbf{M}=\mathrm{M}\cup
\overline{\mathrm{M}}$ and introduce
$$
T_{\mathbf{S},\mathbf{M}}=\raisebox{.7cm}{$\curvearrowright $} 
\hskip -.75cm\prod\limits_{j=-\infty}^{\infty}
T_{j,\mathbf{M}} (\z_0 ^{(-1)^j})\,,\qquad
T_{j,\mathbf{M}}(\z) =\raisebox{.7cm}{$\curvearrowleft $} 
\hskip -.6cm\prod\limits_{\mathbf{m=1}}^{2\mathbf{n}}
L_{j,\mathbf{m}}(\z\z_0^{-(-1)^\mathbf{m}})
\,.
$$
Here
$$
L_{j,\mathbf{m}}(\zeta)
=q^{-\frac 1 2\sigma ^3_j\sigma ^3_\mathbf{m}}
-\z ^2 q^{\frac 1 2\sigma ^3_j\sigma ^3_\mathbf{m}}
-\z (q-q^{-1}) (\sigma ^+_j\sigma ^-_\mathbf{m}+\sigma ^-_j\sigma ^+_\mathbf{m})\,.
$$ 
Note that for $\z_0^{\pm1}=\infty$ the inhomogeneous lattice reduces to two non-interacting homogeneous lattices.

The functional $Z^\mathrm{full}_\mathbf{n}$ is defined by
\begin{align}
Z^\mathrm{full}_\mathbf{n}\Bigl\{q^{2\al S(0)}\mathcal{O}
\Bigr\}=
\frac{\Tr _{\mathbf{S}}\Tr _{\mathbf{M}}\Bigl(T_{\mathbf{S},\mathbf{M}}q^{2\al S(0)}\mathcal{O}\Bigr)}
{\Tr _{\mathbf{S}}\Tr _\mathbf{M}\Bigl(T_{\mathbf{S},\mathbf{M}}q^{2\al S(0)}\Bigr)}
\,.
\label{Zfull}
\end{align}
We set $\kappa=0$ here.
The methods of II,III allow one 
to compute this functional with inhomogeneities in both directions. 
In particular, from II
one concludes that the annihilation operators 
split into two parts:
\begin{align}
&\bb (\z)=\bb ^+(\z)+\bb ^-(\z), \qquad \cb (\z)=\cb ^+(\z)+\cb ^-(\z)\,,\nn\\
&\bb ^{\pm}(\z)=\sum\limits _{p=0}^\infty\(\z^2\z_0^{\mp 2}-1\)^{-p}\bb^\pm_p,
\qquad \cb ^{\pm}(\z)=\sum\limits _{p=0}^\infty\(\z^2\z_0^{\mp 2}-1\)^{-p}\cb^\pm_p\,.
\nn
\end{align}
Let us take $\mathbf{n}=\infty$.
Then using the remark from the previous
section and II,III one obtains:
\begin{align}
Z^\mathrm{full}_\infty \Bigl\{q^{2\al S(0)}\mathcal{O}
\Bigr\}=
\frac 
{\Tr_\mathbf{S}\bigl(e^{\Omega^\mathrm{full}}\(q^{2\al S(0)}\mathcal{O}\)\bigr)}
{\Tr_\mathbf{S}\(q^{2\al S(0)})\)}\,.
\end{align}
We have
\begin{align}
&\Omega^\mathrm{full}=\Omega_{0} +\nabla\Omega \,,\nn
\end{align}
where
\begin{align}
&\Omega_{0}=\Omega_{0}^{++}+\Omega_0^{+-}+\Omega_{0}^{-+}+\Omega_{0}^{--}\,,\nn\\
&\Omega_{0}^{\epsilon\epsilon'}=
\frac{4}{(2\pi i)^2}
\int\limits _{\Gamma_{\epsilon}}
\int\limits _{\Gamma   _{\epsilon'}  }\omega_{0} (\z/\xi,\al)
\cb ^{\epsilon'}(\xi)\bb^{\epsilon} (\z)\frac{d\z^2}{\z ^2}\ \frac{d\xi^2}{\xi ^2}\,,
\nn
\end{align}
and similarly for $\nabla\Omega $.
The contour $\Gamma_{\pm}$ goes anticlockwise around $\z _0^{\pm 2}$.


As it has been said in the introduction, 
ideally we would like to 
start from a non-critical (XYZ or SOS) lattice model
and to obtain the relativistic massive model by
the usual scaling limit near the critical point. 
Since we do not have the necessary formulae to do that, 
we have recourse to the scaling limit of
an inhomogeneous model 
by the procedure of \cite{DDV}.
Here we have an ideal situation from the point of view of QFT.  
Namely, we have the ultraviolet cutoff (lattice), 
the infrared cutoff (a finite number $\mathbf{n}$
of sites in the Matsubara direction),  
and the physical quantities 
(the values of the the functional
$Z^{\mathrm{full}}_\mathbf{n}$ on quasi-local fields) 
are exactly computed with finite cutoffs. 

Like in the homogeneous case, 
let us introduce the step of the lattice $a$, 
and  consider the scaling limit \eqref{scal}: 
$\mathbf{n}\to\infty$,  $a\to 0$, $\mathbf{n}a=2\pi R$ fixed. 
We require further that 
$\z _0^{-1}\to 0$, 
so that 
\begin{align}
M=4 a^{-1}\z _0^{ -1/ \nu}\ \ 
\mathrm{fixed}\,.
\label{mass}
\end{align}
The parameter $M$ is a mass scale which has the meaning of the 
sG soliton mass \cite{DDV}.
The famous formula relating the soliton 
 mass to the dimensional coupling constant $\mub$ 
 \cite{AlZ} reads in our notation as
 \begin{align}
 \mub =\left[\frac {M}{4C}\right]^\nu=\(Ca\)^{-\nu}\z _0^{-1}\,.
 \label{Cmub}
 \end{align}

In this paper we consider a further limit $R\rightarrow\infty$. In that case
the sG partition function is obtained from $Z^{\rm full}_\infty$.

Let us give some explanation at this point. 
The subject of study in \cite{DDV}
is the partition function of the sG model on the cylinder. 
There are two possible approaches to this partition function 
which correspond to two Hamiltonian pictures. 
In the first one, the space direction 
is considered as space and the Matsubara direction as time (space channel).
One has scattering of particles and describes the partition
function by the Thermodynamic Bethe Ansatz (TBA) \cite{TBA}.
This approach has an advantage of dealing with known particle
spectrum and $S$-matrices. 
It also has a disadvantage, 
because as usual in the thermodynamics 
one has to deal with the density matrix, 
which is a complicated object even in integrable cases.

The paper \cite{DDV} uses an alternative picture: 
the Matsubara direction is space, and the space direction is time. 
In this approach, 
the partition function is described by the  maximal 
eigenvalue of the Hamiltonian of the
periodic problem for  the Matsubara direction (Matsubara channel). 
The advantage of this approach is clear: one deals with the 
pure ground state instead of the density matrix. 
The disadvantage is that describing eigenvalues in 
the finite volume is a difficult problem. 
This problem is addressed in \cite{DDV}. 

More precisely, it is proposed in \cite{DDV} to obtain 
the sG partition function as the scaling limit
of $\Tr_\mathbf{S}\Tr_\mathbf{M}(T_{\mathbf{S},\mathbf{M}})$.
As in the present paper,  it is important 
to be able to control the computations starting 
from the lattice and from finite $\mathbf{n}$.
In our opinion, the main achievement
of \cite{DDV} is not in rewriting the Bethe Ansatz equation 
for the Matsubara transfer matrix in the form of 
a non-linear integral equation,
but in extracting 
a main linear part and inverting it. 
The resulting Destri-DeVega equation (DDV)
has several nice features. 
First, it allows the scaling limit and the mass scale
\eqref{mass} appears. 
Second, after the scaling limit the DDV equation clearly
allows   the large $R$ expansion. 
Third, the scattering phase of the sG solitons
appears in the DDV equation in the Matsubara channel. 
The last property
allows the
identification with the space channel. 
In particular, $M$ happens to be equal 
to the mass of soliton. 

 

Now we come to the most important point of this paper. 
We wish to define the creation operators 
appropriate for taking the scaling limit to the sG theory.  
The issue is similar to the one in 
conformal perturbation theory, where  
one needs to prescribe a way how to extend
the descendants in CFT to the perturbed case. 
It is claimed in \cite{alzam,fflzz} that after subtracting the
divergencies these operators are defined uniquely: 
possible finite counterterms can be 
dismissed for dimensional reasons, 
at least in the absence of resonances \cite{alzam,fflzz}. 
The latter condition is satisfied in our case if
$\nu$ and $\al$ are generic.

Let us define
the operators $\bb ^*_{0}(\z)$ and $\cb ^*_{0}(\z)$ by the same formula as in the 
homogeneous case \eqref{bogolub}.
Starting from  these operators  we define
the creation operators 
\begin{align}
\bb ^{ *}_{0}(\z)
\ \ \simeq\hskip -.6cm{}_{{}_{{}_{\scalebox{.6}
{$\z^2\to \z _0^{\pm2}$}}}}
\ 
\bb ^{\pm *}_{0}(\z)=\sum_{p=1}^{\infty}
(\z ^2\z _0^{\mp2}-1)^{p-1}\bb ^{\pm *}_{{0},p}
\nn
\end{align}
and likewise for $\cb ^{*}_{0}(\z)$.  

The operators $\bb^{\pm *}_{0,p}$
$\cb^{\pm *}_{0,p}$ create quasi-local fields.
Notice that $Z^\mathrm{full}_\infty$ is 
defined on the quotient space  
$\mathcal{W}^{(\al)}_\mathrm{quo}$ 
because the $\tb^*$-descendants do not contribute to it.

There are  two sorts of chiral operators, 
one living on the even sublattice 
and the other on the odd sublattice. 
We have to make some combinations which will give
finite answers for the interacting model. 
At the same time, we want this combination to 
correspond to our intuitive idea that
we have to subtract perturbative series. 
Let us explain that the correct combinations are 
given by the following. 
\begin{align}
&\bb ^{+*}(\z)
=e^{-\Omega_0 ^{++}}\bb ^{+*}_{0}(\z)e^{\Omega_0^{++}}
\,,
\quad \cb ^{+*}(\z)
=e^{-\Omega_0 ^{++}}\cb ^{+*}_{0}(\z)e^{\Omega0 ^{++}}\,,
\label{defbcsG}
\\
&\bb ^{-*}(\z)
=e^{-\Omega_0^{--}}\bb ^{-*}_0(\z)e^{\Omega_0 ^{--}}
\,,
\quad \cb ^{-*}(\z)
=e^{-\Omega_0^{--}}\cb ^{-*}_0(\z)e^{\Omega_0 ^{--}}\,.
\nn
\end{align}
Corresponding creation operators which create the
quasi-local operators are defined by
$$
\bb ^{ \pm *}(\z)
\ \ \simeq\hskip -.6cm{}_{{}_{{}_{\scalebox{.6}
{$\z^2\to \z _0^{\pm2}$}}}}
\sum_{p=1}^{\infty}
(\z ^2\z _0^{\mp2}-1)^{p-1}\bb ^{\pm *}_{p}
\nn
$$

In what follows we shall be interested only in the case of an equal 
number of $\bb^{+*}$ and $\cb^{+*}$. 
Let us examine how the descendants of this form 
depend on $\z_0$, taking the simplest case
\begin{align*}
&\bb_k^{+*}\ \cb_l^{+*}\ \bb_r^{-*}\ \cb_s^{-*}(q^{2\al S(0)})
=e^{-\Omega_0^{++}-\Omega_0^{--}}
\bb_{0,k}^{+*}\ 
\cb_{0,l}^{+*}\ 
\bb_{0,r}^{-*}\ 
\cb_{0,s}^{-*}(q^{2\al S(0)})\,.
\end{align*}
It is easy to see from the definition in II that 
the second factor in the right hand side 
is a rational function of $\z_0^2$ regular at $\infty$, 
\begin{align}
&\bb _{0,k}
^{+*}\ \cb
_{0,l}^{+*}\ \bb
_{0,r} ^{-*}\ \cb 
_{0,s}^{-*}(q^{2\al S(0)})
=
\(\mathcal{O}_0+\z ^{-2}_0
\mathcal{O}_1+\z_0 ^{-4}\mathcal{O}_2+\cdots\)
q^{2\al S(0)}\,.
\label{decomp}
\end{align}
The rationality holds true after application of 
$e^{-\Omega_{0} ^{++}-\Omega_{0} ^{--}}$. 
The expansion \eqref{decomp} looks as a perturbative series 
for the action \eqref{action}, 
because in the scaling limit $\z _0^{-2}$ 
comes accompanied by $a ^{-2\nu}$,
and $\z _0^{-2}a^{-2\nu}$ has the dimension 
of $(\mathrm{mass})^{2\nu}$ 
i.e. that of  $\mub ^2$  (see \eqref{mass}).
Notice that the property \eqref{decomp} would be spoiled if we apply 
$e^{-\Omega_{0} ^{+-}-\Omega_{0} ^{-+}}$ as well,  
because it will pick up $\omega_{0}(\z,\al)$ 
near $\z^2=\z_0^{\pm2}$,
and the function 
$\omega_{0} (\z,\al)$ has asymptotics at $\z\to\infty$
containing both $\z ^{\al-2m}$ and $\z ^{-\frac{2n-1}\nu}$.

On the descendants created by \eqref{defbcsG}, 
the value of $Z^{\mathrm{full}}_\infty$ 
remains finite in the scaling limit.  
So we conclude that they create renormalised local operators. 
We cannot say anything about the finite renormalisation, 
but it can be taken care of by a dimensional consideration
(see below).

We conjecture that the following limits exist
\begin{align}
&\half\lim\limits_{\mathrm{scaling}}\bb ^{+*}(\z)  
\ \ \ \simeq\hskip -.6cm{}_{{}_{{}_{\scalebox{.7}{$\z^2\to \infty$}}}}\ \betab^*(\mub\z)+\bar{\betab}^*_\mathrm{screen}
(\z/\mub)\,,\label{sclim}
\\&
\half\lim\limits_{\mathrm{scaling}}\cb ^{+*}(\z) \
\ \ \ \simeq\hskip -.6cm{}_{{}_{{}_{\scalebox{.7}{$\z^2\to \infty$}}}}\ 
\gammab^*(\mub\z)+\bar{\gammab}^*_\mathrm{screen}
(\z/\mub)\,,\nn\\
&\half\lim\limits_{\mathrm{scaling}}\bb ^{-*}(\z)\ 
\ \ \ \simeq\hskip -.6cm{}_{{}_{{}_{\scalebox{.7}{$\z^2\to 0$}}}}\ 
\ \bar{\betab}^*(\z/\mub)+{\betab}^*_\mathrm{screen}
(\mub \z)\,,\nn\\
&\half\lim\limits_{\mathrm{scaling}}\cb ^{-*}(\z)\ 
\ \ \ \ \simeq\hskip -.6cm{}_{{}_{{}_{\scalebox{.7}{$\z^2\to 0$}}}}\ 
\ \bar{\gammab}^*(\z/\mub)+\gammab^*_\mathrm{screen}
(\mub \z)\,,\nn
\end{align}
where by $\lim_{\mathrm{scaling}}$
the scaling \eqref{scal}, \eqref{mass} and $R\rightarrow\infty$ are implied.

In these formulae we denote by the same letters the operators in the sG model as they were denoted in the CFT.
We shall not consider the screening operators,while
we use the coefficients $\betab^*_{2j-1}$ and so forth, defined as in
\eqref{betgam} and \eqref{betgambar}, to consider the descendants
\begin{align}\label{DESCEN}
\bar{\betab} ^{*}_{\bar{I}^+}\bar{\gammab} ^{*}_{\bar{I}^-}
{\betab} ^{*}_{{I}^+}{\gammab} ^{*}_{{I}^-}\Phi _\al (0)\,.
\end{align}
This is the field in the interacting model which goes to corresponding
descendant in the conformal limit $\mub\to 0$ , and which does not
develop finite counterterms. The latter are forbidden by dimensional consideration.
So, this is exactly the definition which we were supposed to use from the very beginning. 
Notice that the appearance of $\mub$ in the formulae
\eqref{sclim}
is a consequence of consistency with the conformal limit due to
\eqref{Cmub}.

Now it is easy to compute the normalised
vacuum expectation value of the descendant \eqref{DESCEN} for the sG model.
It is obtained by taking the scaling limit of $Z^\mathrm{full}_\infty$ and
computing the asymptotics of $\omega_{0} (\z,\al)$ for $\z \to\infty$ and $\z\to 0$ which is done simply by
summing up the residues in appropriate half-planes:
\begin{align*}
\omega_{0}(\z,\al)\simeq\frac i\nu\sum_{n\geq1}\z^{-\frac{2n-1}\nu}\cot\frac{\pi}{2\nu}\(2n-1 +\nu\al\)
+i\sum_{n\geq1}\z^{\al-2n}\tan\frac{\pi\nu}2\(\al-2n\).
\end{align*}
The first part corresponds to the expectation values of operators created by the coefficients
of the expansions \eqref{betgam}, \eqref{betgambar}.
These are the expectation values in which we are interested in this paper.
The second part of the asymptotics corresponds to the operators 
\eqref{scr1}, \eqref{screenbar}
At present we do not know the meaning of their expectation
values, but we hope to return to them in the future.

Introducing the multi-indices as described in the introduction
we obtain:
\begin{align}
&\frac
{\langle  
\bar{\betab} ^{*}_{\bar{I}^+}\bar{\gammab} ^{*}_{\bar{I}^-}
{\betab} ^{*}_{{I}^+}{\gammab} ^{*}_{{I}^-}\Phi _\al (0)
\rangle_\mathrm{sG}}
{\langle  
\Phi _\al (0)
\rangle_\mathrm{sG}}=\delta _{\bar{I}^-,I^+}\delta _{\bar{I}^+,I^-}
(-1)^{\#(I^+)}
\(\frac{ i}{\nu}\)^
{\#(I^+)+\#(I^-)}\label{main0}\\
&\times \mub^{\frac{2}{\nu}(|I^+|+|I^-|)}\prod\limits _{2n-1\in I^+}\cot
\textstyle{\frac \pi 2 }(
\textstyle{\frac{2n-1}{\nu}}+\al)
\prod\limits _{2n-1\in I^-}
\cot
\textstyle{\frac \pi 2 } (
\textstyle{\frac{2n-1}{\nu}}-\al)
\,.
\nn
\end{align}
Now we have to recall the definition in IV:
\begin{align}
\betab^*_{2n-1}=D_{2n-1}(\al)\betab^{\mathrm{CFT}*}_{2n-1}\,,
\quad \gammab^*_{2n-1}=D_{2n-1}(2-\al)\gammab^{\mathrm{CFT}*}_{2n-1}\,,
\end{align}
where 
$$D_{2n-1}(\al)=
\frac {1}{\sqrt{i\nu}}\ G^{2n-1}
\ \frac{\Gamma\bigl(\frac{\alpha}{2}+\frac 1{2\nu}(2n-1)\bigr)}
{(n-1)!\ \Gamma\bigl(\frac{\alpha}{2}+\frac{(1-\nu)
}{2\nu}(2n-1)
\bigr)}\,,$$
with
$$G=\Gamma (\nu)^{-1/\nu}\sqrt{1-\nu}\,.$$
Similarly we get for the second chirality
\begin{align}
\bar{\betab}^*_{2n-1}=D_{2n-1}(2-\al)\bar{\betab}^{\mathrm{CFT}*}_{2n-1}\,,
\quad\bar{ \gammab}^*_{2n-1}=D_{2n-1}(\al)
\bar{\gammab}^{\mathrm{CFT}*}_{2n-1}\,. 
\end{align}
The main formula \eqref{main} follows immediately.

Before concluding this paper, 
let us say that in principle our approach
can be applied
to the computation of one-point functions
of descendants for finite radius in the Matsubara direction
(finite temperature, in other words). 
However, this would require a detailed study of 
the DDV equation
and the equations for the function $\omega_R$ 
for the sG model in finite volume.

\bigskip

{\it Acknowledgements.}\quad
Research of MJ is supported by the Grant-in-Aid for Scientific 
Research B-20340027. 
Research of 
TM is supported by the Grant-in-Aid for Scientific Research
B-22340017.
Research of FS is supported by  RFBR-CNRS grant 09-02-93106.

Taking this occasion, MJ and FS wish to 
offer TM  congratulations on his sixtieth birthday. 
\bigskip


\end{document}